\documentclass[%
 reprint,
 amsmath,amssymb,
pra,
]{revtex4-1}

\usepackage{graphicx}
\usepackage{dcolumn}
\usepackage{bm}

\begin{document}


\title{Cell division and death inhibit glassy behaviour of confluent tissues}

\author{D. A. Matoz Fernandez}
\affiliation{%
 ~Universit\'{e} Grenoble Alpes, LIPHY, F-38000 Grenoble.
}%
\email{daniel-alejandro.matoz-fernandez@univ-grenoble-alpes.fr}
\altaffiliation{
New affiliation: Division of Molecular Microbiology, College of Life Sciences, University of Dundee, Dundee DD1 5EH. damatozfernandez@dundee.ac.uk
}

\author{Kirsten Martens}
\affiliation{%
 ~Universit\'{e} Grenoble Alpes, LIPHY, F-38000 Grenoble.
}%
\author{Rastko Sknepnek}
\affiliation{%
School of Science and Engineering and School of Life Sciences, University of Dundee, Dundee DD1 4HN, United Kingdom.
}%
\author{J. L. Barrat}
\affiliation{%
 ~Universit\'{e} Grenoble Alpes, LIPHY, F-38000 Grenoble.
}%
\author{Silke Henkes}
\affiliation{
ICSMB, Department of Physics, University of Aberdeen, Aberdeen AB24 3UE, United Kingdom.
}%


\begin{abstract}
We investigate the effects of cell division and apopotosis on collective dynamics in two-dimensional epithelial tissues. Our model includes three key ingredients observed across many epithelia, namely cell-cell adhesion, cell death and a cell division process that depends on the surrounding environment. We show a rich non-equilibrium phase diagram depending on the ratio of cell death to cell division and on the adhesion strength. For large apopotosis rates, cells die out and the tissue disintegrates. As the death rate decreases, however, we show, consecutively, the existence of a gas-like phase, a gel-like phase, and a dense confluent (tissue) phase. Most striking is the observation that the tissue is self-melting through its own internal activity, ruling out the existence of any glassy phase.
\end{abstract}
\maketitle


\section{\label{sec:Intro}Introduction}

Simple epithelial tissues consist of a single layer of tightly connected cells. Especially during development, epithelial cells grow, divide and move, leading to a dynamic reorganisation of the entire tissue. This process is regulated by a complex set of chemical and mechanical signalling pathways~\cite{lecuit2011force, janmey2011mechanisms, mammoto2010mechanical, tenney2009stem} that control cell shapes and cell-cell contacts. How the regulation of cell-cell interactions is transmitted to the tissue-level organisation is still a topic of active research. 
Mechanical signalling, \emph{i.e.}, a set of processes that control the cell response to mechanical stimuli in the form of externally applied or internally generated forces, is at present only partly understood.~\cite{janmey2011mechanisms}
One well-known example of mechanics-influenced regulation is the density-dependent inhibition of proliferation in cell monolayers.~\cite{Martz1972, alberts2007cell} A hallmark of cancerous tissues is the absence of this regulation, leading to uncontrolled tumour growth.
Perturbations in the mechanical sensing of cells have been reported to 
be relevant in several diseases such as osteoporosis and atherosclerosis.~\cite{jacobs2013introduction} Breast cancer,~\cite{paszek2005tensional} cardiovascular~\cite{engler2008embryonic} and liver diseases~\cite{li2007transforming} as well as renal glomerular disease~\cite{tandon2007hiv} are all known to be accompanied 
by significant changes in the mechanical properties of relevant tissues. 

Recent advances in microscopy techniques and powerful algorithms for 
automated cell tracking have enabled studies of collective cell 
migration for large cell numbers, over extended periods of time and 
with high spatial resolution, both \emph{in vitro} and \emph{in vivo}. 
Traction force microscopy~\cite{harris1980silicone} measurements revealed that the collective motion of epithelial cell layers is far more complex than previously believed.~\cite{angelini2011glass,tambe2011collective,trepat2011plithotaxis} Homogeneous cell sheets behave as a supercooled fluid at long time scales and as a glass at short time scales, showing large spatial fluctuations of the inter-cellular forces. These fluctuations cannot be pinpointed to a specific cell but extend over regions spanning several cells.~\cite{Szabo2006, Trepat2009, sadati2013collective} They strongly resemble the fluctuations observed in supercooled colloidal and molecular liquids approaching the glass transition~\cite{angelini2011glass} with evidence of dynamical heterogeneity, a hallmark of glassy dynamics that has been extensively studied in soft condensed matter physics.

In spite of the many interesting similarities to soft glasses, cell 
sheets viewed as active materials constitute a new class of 
non-equilibrium system in which the interplay between activity, long 
range elasticity and cell interactions give rise to novel phases with 
unusual structural, dynamical and mechanical 
properties.~\cite{Fily2012, Fily2014, Bi2016} Many recent works have shown that cell activity, for example in the form of self-propulsion, has the capability to fluidise a confluent tissue, but only above a critical level of activity.~\cite{Fily2014, Berthier2014, Bi2016, Dasgupta2016} At low enough activity all of these works report the existence of a glassy phase where cell diffusion ceases.

In contrast, in this paper we show that the simple presence of any finite rate of cell division and death completely destroys the glassy dynamics of the tissue. In agreement with Ranft \emph{et}. al ~\cite{Ranft2010}, we report that cell division and apopotosis always fluidises the confluent tissue. 
To systematically explore the effect of cell division and cell death as an active driver, we introduce a minimal particle-based model based on simplified cell and division dynamics. This allows us to fully explore the phase space of the model and enumerate its phases, from gaseous to  gel-like and eventually confluent, as a function of the relative death to division ratio (Figure \ref{fig:2}). We carefully characterise the lower-density transitions (absorbing to gaseous, phase separated to gel-like) to produce a phase diagram (Figure \ref{fig:5}). In the confluent phase, we show the self-melting effect of a range of division and death rates, and their scaling limits (Figure \ref{fig:divide_dense}). Finally, we compare division and death dynamics to active self-propulsion dynamics and show that at the long time scales relevant to glassy dynamics, the effect of division always dominates (Figure \ref{fig:divide_and_active}).

\section{Model}

Cell shape is known to play an important role in tissue organisation, 
and it is controlled by a complex set of signalling 
pathways.~\cite{lecuit2007cell} Despite its complexity, a remarkable amount of information about collective behaviour at scales exceeding the size of a single cell can be gained from effective models that treat cells as soft elastic objects.~\cite{zimmermann2016contact} More generally, particle based tissue models have been successfully applied to a wide range of systems (for a complete review see Drasdo \textit{et al.}~and the references within~\cite{Drasdo2007}). In this study we take a similar approach and consider a model where the cells are represented by soft spheres of radius $b_i$. The tissue consists of a collection of $N$ such spheres with radii $b_{i}$ uniformly distributed in the range of $0.85$ to $1.15$. 

\begin{figure}[h!]
\centering
\includegraphics[width=0.99\columnwidth]{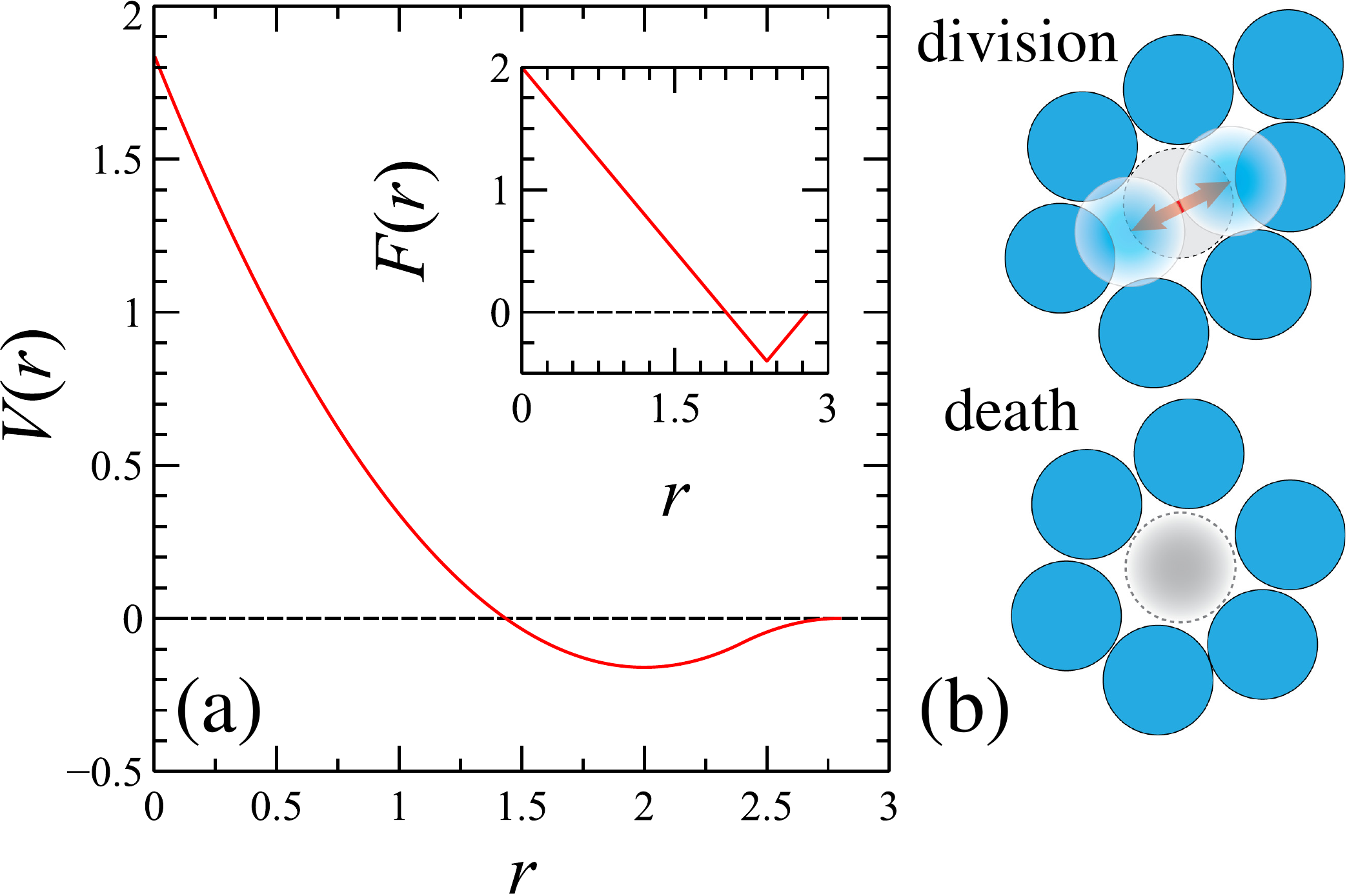}
\caption{(a) Interparticle potential $V(r)$, for $k = 1$ and $\epsilon = 0.2$. The elastic force is shown as inset. (b) Schematic illustration of the cell division and cell death dynamics.}
\label{fig:1}
\end{figure}

We model the contact forces between two cells $i$ and $j$ through a pair potential that includes short range repulsion to mimic volume exclusion, together with short range adhesion (see Figure ~\ref{fig:1}(a)).~\cite{Szabo2006, Drasdo2007} The potential is given by 
\begin{equation}
\label{eq:1}
V(r_{ij}) = 
\begin{cases}
\frac{1}{2} k b_{ij}^{2} \left[\left(\frac{r_{ij}}{b_{ij}}-1\right)^2 - \epsilon^2\right]&\mbox{if } \frac{r_{ij}}{b_{ij}}-1  \leq \epsilon  \vspace{0.1cm}\\
-\frac{1}{2} k b_{ij}^{2} \left(\frac{r_{ij}}{b_{ij}}-1-2\epsilon\right)^2 &\mbox{if } \epsilon < \frac{r_{ij}}{b_{ij}}-1 \leq 2\epsilon,
\end{cases}
 \end{equation}
where $k$ is the stiffness constant, ${b_{ij} = b_{i} + b_{j}}$ is the sum of the particle radii, and $(b_{ij}\epsilon)$ is the adhesive force strength.

In accordance with micron-size scales for cell diameters, we neglect inertia effects and model the dynamics of the cell positions $\mathbf{r}_{i}(t)$ as fully overdamped~\cite{Henkes2011}
 \begin{equation}
 \label{eq:2}
\partial_t \mathbf{r}_{i}(t) =  \mu \mathbf{F}_{i},
\end{equation}
where $\mu$ is the inverse friction coefficient and ${\mathbf{F}_{i} = \sum_{j\neq i}  \mathbf{F}_{ij}}$ is the total force acting on particle $i$ exerted by its neighbours. 

The only source of activity in the system is cell division and 
apoptosis, as schematically drawn in Figure \ref{fig:1}(b). Apoptosis is included by removing cells randomly at constant rate $a$. Note that this simplified approach can also model other removal mechanisms, such as sheet extrusion or ingression from the sheet into other tissues. Motivated by the well-known density-dependent inhibition of proliferation in cell monolayers,~\cite{Martz1972, alberts2007cell} we model cell division as a density dependent mechanism with a division rate
\begin{equation}
\label{eq:3}
d = d_{0}\left(1 - \frac{z}{z_{max}}\right),
\end{equation}
where $d_{0}$ is the division rate amplitude, $z$ is the number of contact neighbours of the particle and $z_{max}$ is number of contact neighbours at which division ceases in the system.
We fix the maximum value of nearest neighbours to $z_{max} = 6$, 
\emph{i.e.} a full ring of nearest neighbours. Taking rearrangements into account, this allows for the neighbour distribution with mean $6$ typical of a two-dimensional confluent tissue,~\cite{Sandersius2011,Gibson2006} see Figure \ref{fig:1}(b).
We replace the cell by the new mother-daughter pair located on top of each other, and then linearly fade in their mutual potential $V_{ij}$, therefore preventing jumps in the local forces.

Our model contains two microscopic time scales: the elastic interaction time scale ${\tau_{\text{el}}=(\mu k)^{-1}}$ and a much longer time scale introduced by the active division process $\tau_{\text{a}}=(d_{0})^{-1}$.  We fix the simulation time unit by setting $\mu = k = 1$. 
Then the phase space can be explored varying only three control 
parameters: $(1)$ the ratio of apoptosis to division rate, $a/d_0$, 
$(2)$ the ratio of attraction to repulsion $\epsilon$. (3) Furthermore, 
we have established that the homeostatic properties of the system 
(density, pressure, contact number) do not depend on $d_0$ (see SI, 
section A). We study the dynamics of the model in a square box of size 
$L = 120$ with periodic boundary conditions to mimic the bulk dynamics 
of the tissue. Depending on final density, this is equivalent to 
$N=2000-10000$ particles. The simulations were carried out using both a 
C$++$ GPU-parallel Molecular Dynamics code (see SI, 
section D), and the multi-purpose active matter simulation code SAMoS 
(Soft Active Matter on Surfaces).~\cite{SamosPaper}\\

\section{Results and Discussion}

\begin{figure*}
\centering
\includegraphics[height=7cm,keepaspectratio]{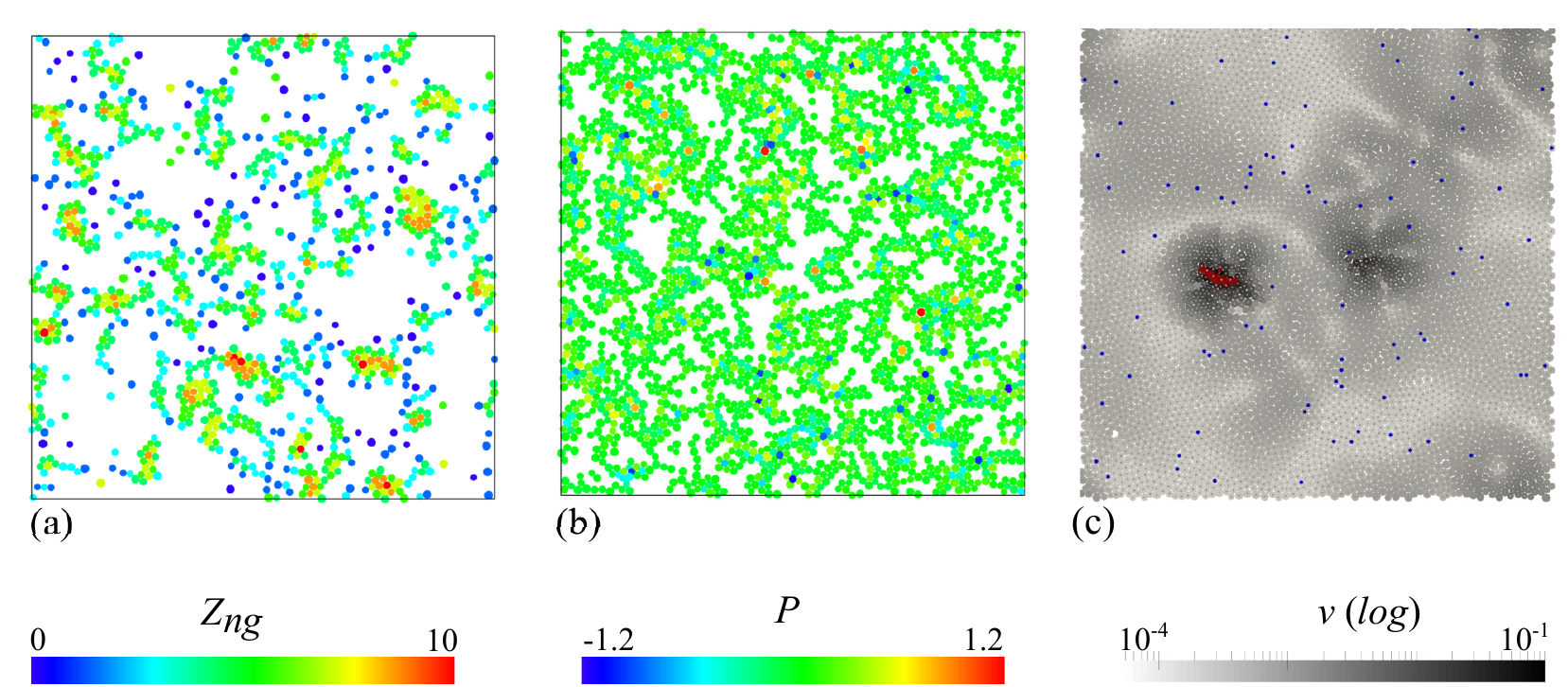}
\caption{Snapshots of the system in different parts of the phase space for $\epsilon = 0.15$. (a) Gas-like phase at apoptosis rate just within the stable region ($a/d_0 = 0.5$, $d_0 = 10^{-2}$); (b) phase separated state consisting of a percolated cluster surrounded by a gas of cells ($a/d_0 = 0.3$, $d_0 = 10^{-2}$);  (c) self-melting dynamics in a confluent system ($a/d_0=3\times10^{-3}$, $d_0=3\times10^{-3}$). Particles are coloured according to: (a) contact number, (b) local virial pressure and (c) velocity magnitude (log scale). Note that the tracer particles used for the mean-square displacement and self-intermediate function calculation are shown in blue.}
\label{fig:2}
\end{figure*}

To study the interplay between activity and adhesion, we explore the phase space of $a/d_0$ and $\epsilon$.

\begin{figure}[h!]
\centering
\includegraphics[height=7cm,keepaspectratio]{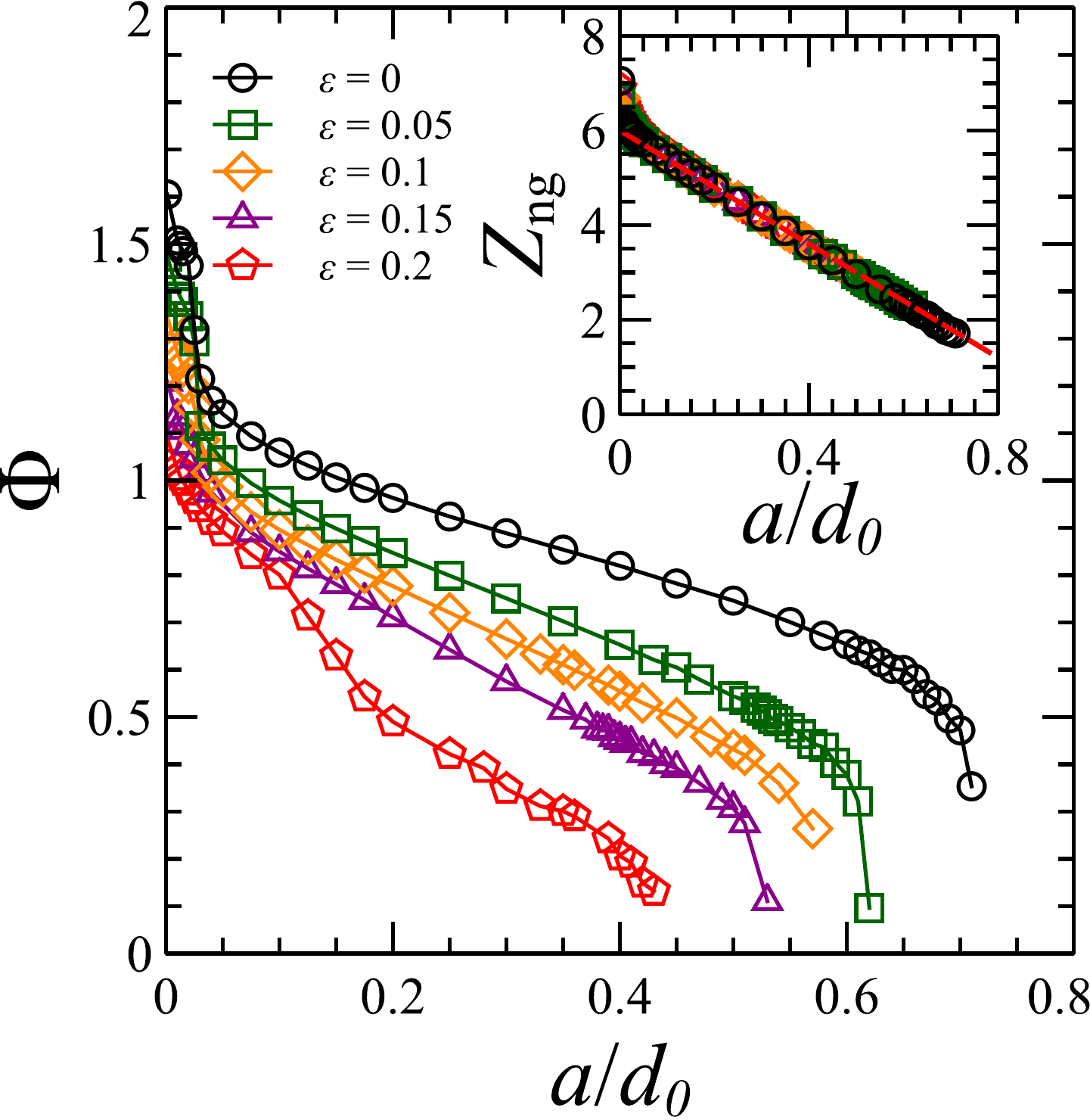}
\caption{Packing fraction $\Phi$ and mean contact number $Z_{\text{ng}}$ (inset) as a function of the ratio of apoptosis to division rate, $a/d_0$.  The mean field line, ${z_{\text{MF}}=z_{\text{max}}(1-a/d_0)}$, is indicated by red dashed lines.  Different  symbols and colours (online) correspond to different attraction forces $\epsilon$. For this figure we used $d_0 = 10^{-2}$.}
\label{fig:3}
\end{figure}

\begin{figure}[h!]
\centering
\includegraphics[height=7cm,keepaspectratio]{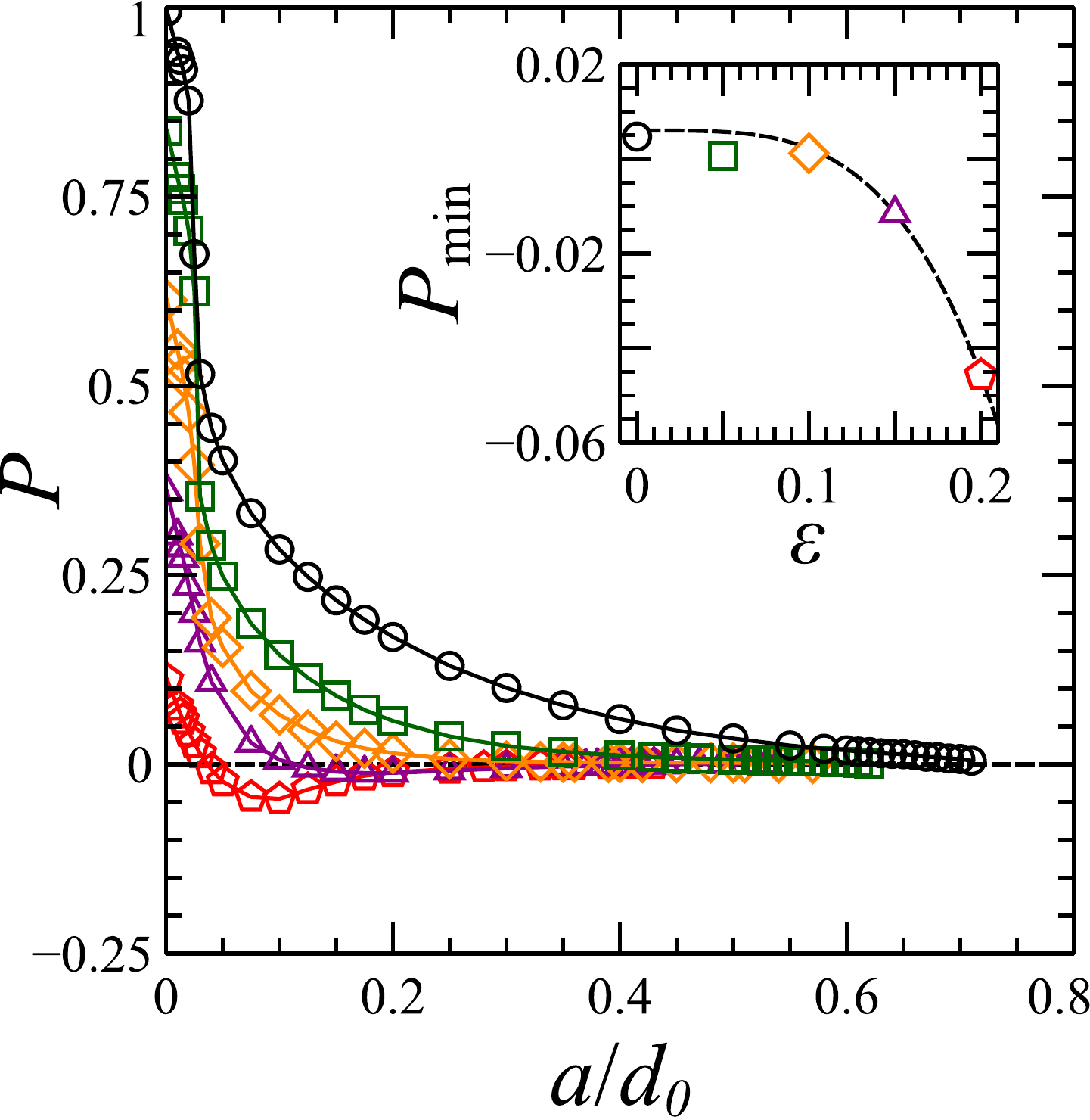}
\caption{Virial pressure $P$ as function of the ratio of apoptosis to division rate, $a/d_0$. The negative pressure region is
indicative of a gel-like phase and the inset shows the minimum pressure, $P_\mathrm{min}$ for different values of $\epsilon$. The line is a guide for the eye. Symbols and colours are the same as in Fig. \ref{fig:3}.}
 \label{fig:4}
\end{figure}

We monitor the state of the system by following the packing fraction ${\Phi = \sum_{i} \pi b_{i}^{2}/L^{2}}$, the number of contact neighbours $Z_{ng}$ and the virial pressure ${P = \sum_i  \mathbf{r}_{i} \cdot \mathbf{F}_{i}/2V}$. The corresponding results are shown in Figures \ref{fig:3} and \ref{fig:4}. 

At high apoptosis rates $a/d_0 \lesssim 1$, the system is unable to reach a steady state at non-zero density, \emph{i.e.} the colony dies out. 
We find an $\epsilon$-dependent critical $a/d_0$ where the cell division is first able to balance cell death and the system reaches a gas-like state (Fig. \ref{fig:2}(a)). Since all of the values are below the expected threshold of stability, $a/d_0=1$, it is clear that collective effects play a role. In the steady state, the rate of loss of particles and the actual division rate balance each other, \emph{i.e.} $\langle d \rangle = a$, where the average takes local correlations into account. 
Intuitively, we can derive the following mean-field scaling for the contact number,
\begin{equation} 
z_{\text{MF}}=z_{\text{max}}(1-a/d_0).
\end{equation}
As shown in the inset to Fig. \ref{fig:3}, the $z-a/d_0$ curves for all $\epsilon$ collapse, and deviations from the linear scaling occur only at the lowest $a/d_0$.

What sets the critical $a/d_0$ value remains an open question. If we extend the mean-field argument to the mean density,  ${\Phi_{\text{MF}}=\Phi_{\text{max}}(1-a/d_0)}$, where $\Phi_{\text{max}}$ is the packing fraction of a system with $\langle z \rangle = z_{\text{max}}$, we obtain a linear scaling that is consistent with much of the intermediate $a/d_0$ range. However, this argument overestimates $\Phi$ when $\epsilon$ is increased.

\begin{figure}[t]
\centering
\includegraphics[height=7cm,keepaspectratio]{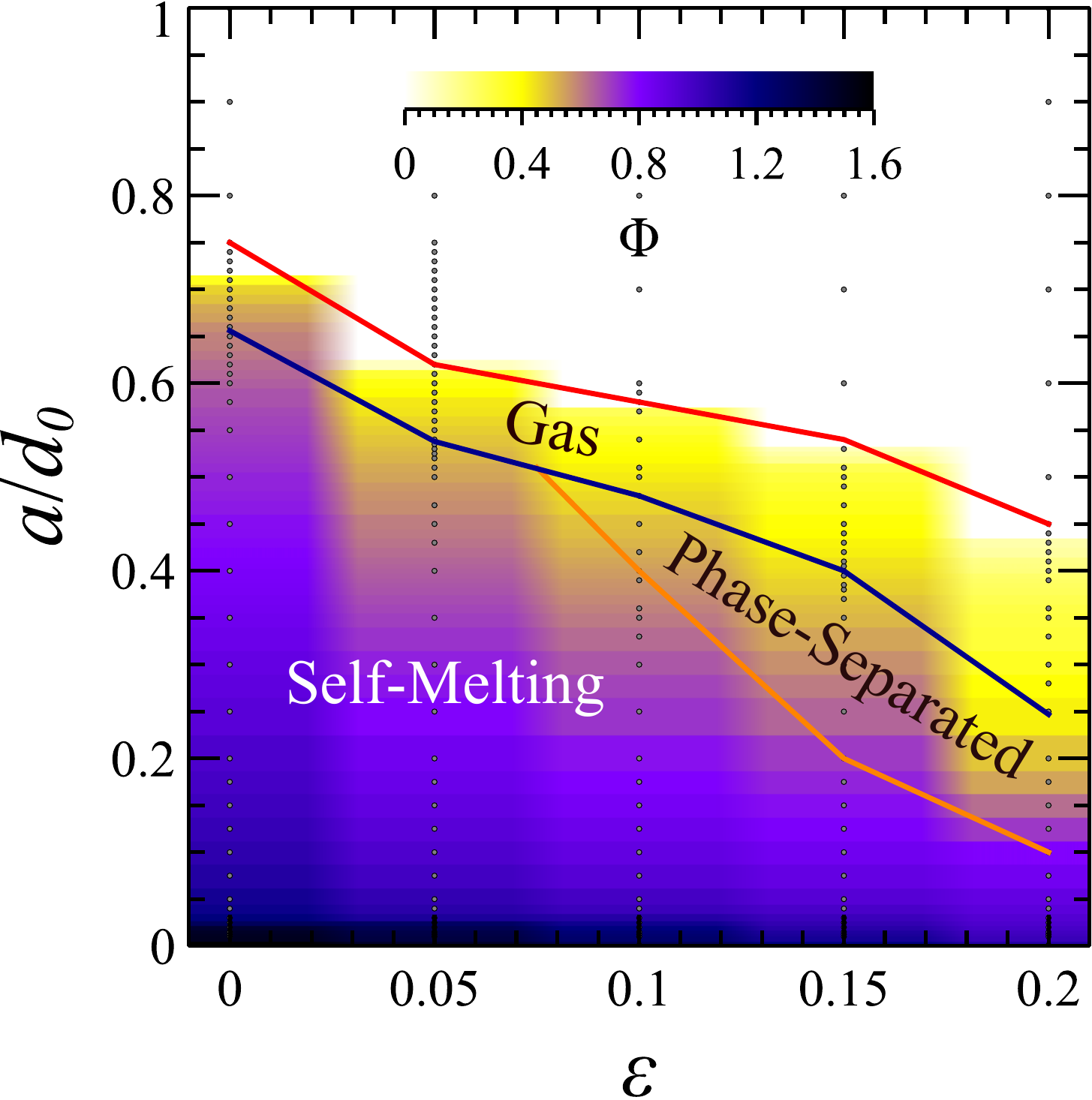}
\caption{ Phase diagram of the system as a function of the adhesion 
force characterised by $\epsilon$ and the activity characterised by $a/d_0$. The red line represent the (numerically estimated) first passage line between an absorbing state and the clustering gas phase. The blue line corresponds to the percolation transition in the system which separates the gas from the percolating cluster phase and the liquid. Finally, the orange line denotes the cluster-liquid transition. The color map shows the value of the packing fraction, $\Phi$.}
 \label{fig:5}
\end{figure}

Clustering is observed at $\epsilon=0$ in the absence of any adhesion 
force, simply due to the fact that cell divisions create new cells nearby.~\cite{Houchmandzadeh2002} Spatial heterogeneities lower the effective division rate since the typical number of neighbours increases, and hence the critical apoptosis rate also decreases. 
As we increase the adhesion force, we observe even stronger spatial heterogeneities and so the effective local division rate decreases more strongly, due to the contact number in the clusters reaching $z_{\text{max}}$. We predict a decrease of the critical $a/d_0$ with $\epsilon$, consistent with the numerical results in Fig.~\ref{fig:5}.
The actual lowest achievable $a/d_0$ is in fact set by a first passage problem: no colony can recover once all cells have died. It is important to note that the finite size has a crucial effect in this situation. Further work is needed to explore this effect in more detail (see SI, section B).

\begin{figure}[t]
\centering
\includegraphics[height=7cm,keepaspectratio]{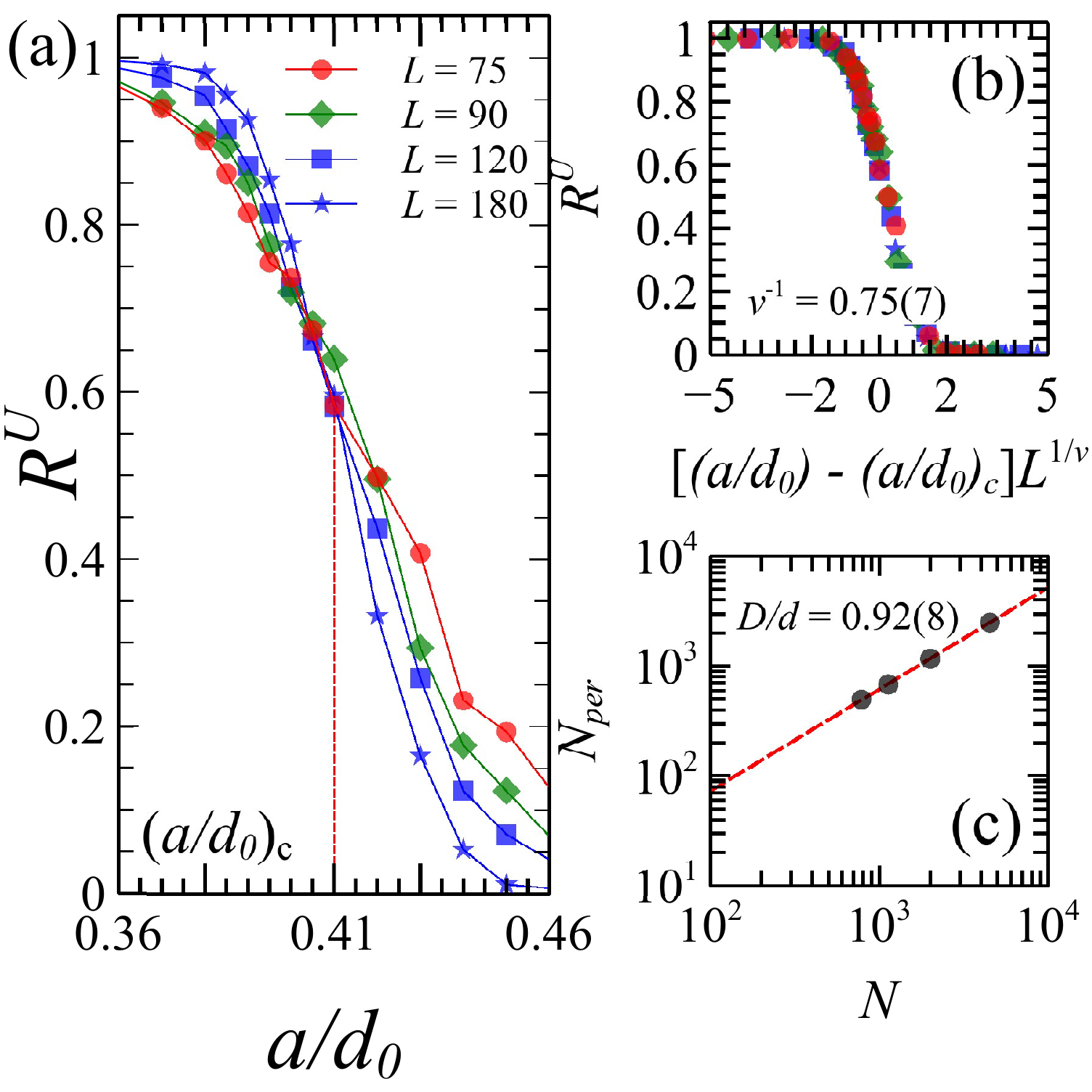}
\caption{(a) Fraction of percolating systems $R^{U} = R^{U}_{L}(a/d_0)$ as function of the $a/d_0$ ratio for the adhesive force $\epsilon = 0.15$. (b) Data collapsing of the fraction of percolating system as function of the reduce $\left[(a/d_0) - (a/d_0)_c\right]L^{1/\nu}$, the critical exponent $\nu = 0.75$. (c) Size of the percolating cluster, $N_{per}$ at the critical point versus the mean number of particles $N$, in a log-log plot the slope is $D/2$ where $D$ is the fractal dimension of the system.}
 \label{fig:percolation}
\end{figure}

In addition, decreasing $a/d_0$ from its critical value causes a rapid 
increase in the density, leading to a gel-like percolated structure 
(Fig. \ref{fig:2}(b)). 
{We investigate the percolation threshold by using the probability of finding a system of size $L$ that percolates in any direction at a given $a/d_0$ ratio, $R^{U} = R^{U}_{L}(a/d_0)$ \cite{Yonezawa1989}. Using finite-size scaling theory, \cite{Binder1997} we can obtain the percolation threshold for each value of $\epsilon$, giving rise to the blue transition line in the phase diagram Fig.~\ref{fig:5}. The results for $\epsilon = 0.15$ are shown in \ref{fig:percolation}(a), and the data collapse of $R^{U}$ in Fig. \ref{fig:percolation}(b) reveals that the critical exponent $\nu$ characerising the divergence of the correlation length observed in our system is consistent with ordinary random percolation.~\cite{stauffer1994introduction}}

{In addition, we measure the fractal dimension by plotting the size of the percolation cluster, $N_{per}$ at at the critical point versus the size of the system $N$, see Fig.\ref{fig:percolation}(c). The value $D \approx 2$ is consistent with independent measures using the structure factor (see Fig. \ref{fig:6}c). The presence of space-filling clusters can be understood from our division process: should any cluster with $D<2$ appear, its mean contact number $z$ will fall significantly below $z_{\text{max}}$, leading to growth to fill in the holes.}

Depending on the strength of the adhesion force $\epsilon$, the final confluent tissue state, Fig. \ref{fig:2}(c) is reached either directly or through phase separation mechanism where a gel-like structure appears in the system. In section \ref{Self-melting} we discuss the self-melting phase in more detail, and the gel phase in section \ref{sec:gel}. Additionally, for large attraction $\epsilon>0.2$, the central soft core repulsion can be overcome, and the system aggregates into an unphysical series of clumps.

We have constructed the $(\epsilon, a/d_{0})$ phase diagram shown in Fig. \ref{fig:5} using the results discussed above, together with the methods described in the followings subsections. 

\begin{figure}[t]
\centering
\includegraphics[width=0.99\columnwidth]{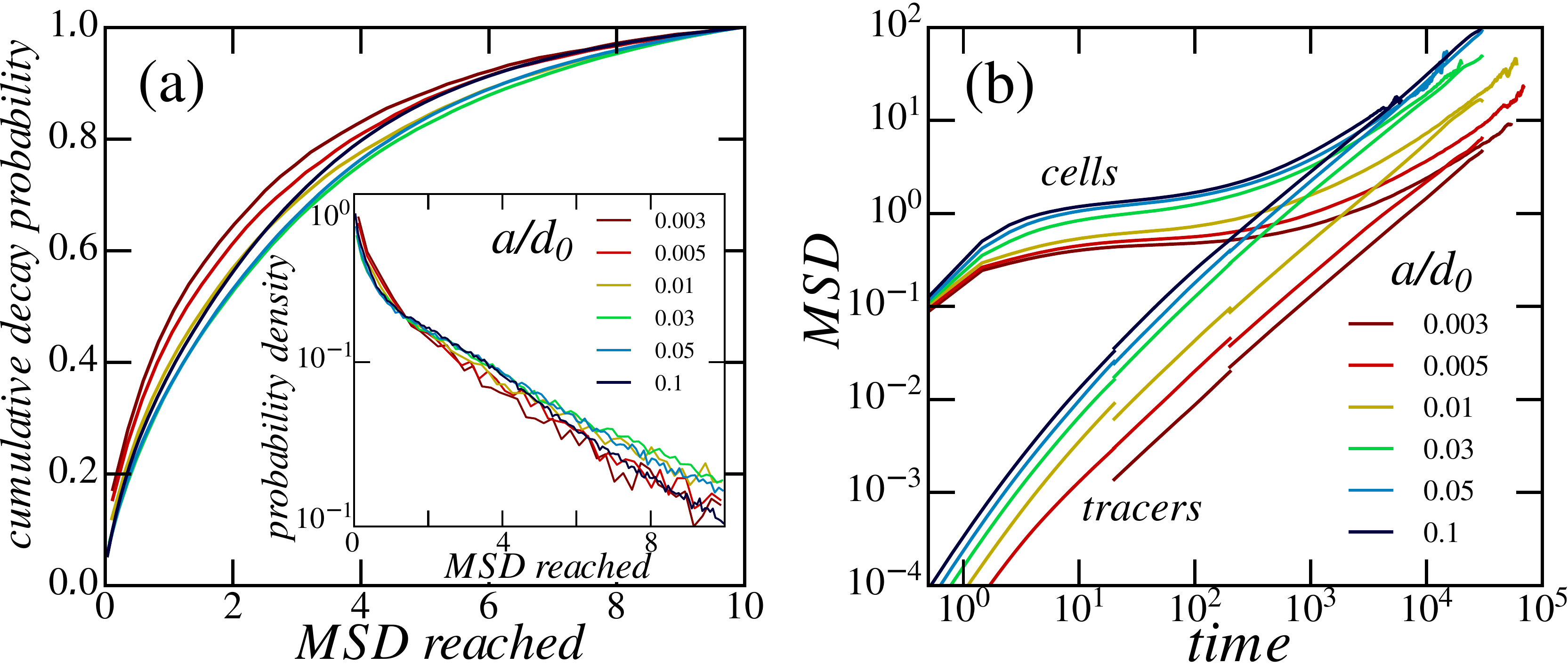}
\caption{{{Mean square displacement of cells and tracers. (a) Probability for cells to decay before reaching a given MSD, for different $a/d_0$. Inset: Probability to reach a certain MSD value. (b) MSD as a function of time for cells since division (top curves) and for tracer particles (bottom curves). All curves are at $d_0=0.01$.}} }
 \label{fig:msd_cells}
\end{figure}

\begin{figure*}[t]
\centering
\includegraphics[width= 0.8\textwidth,clip]{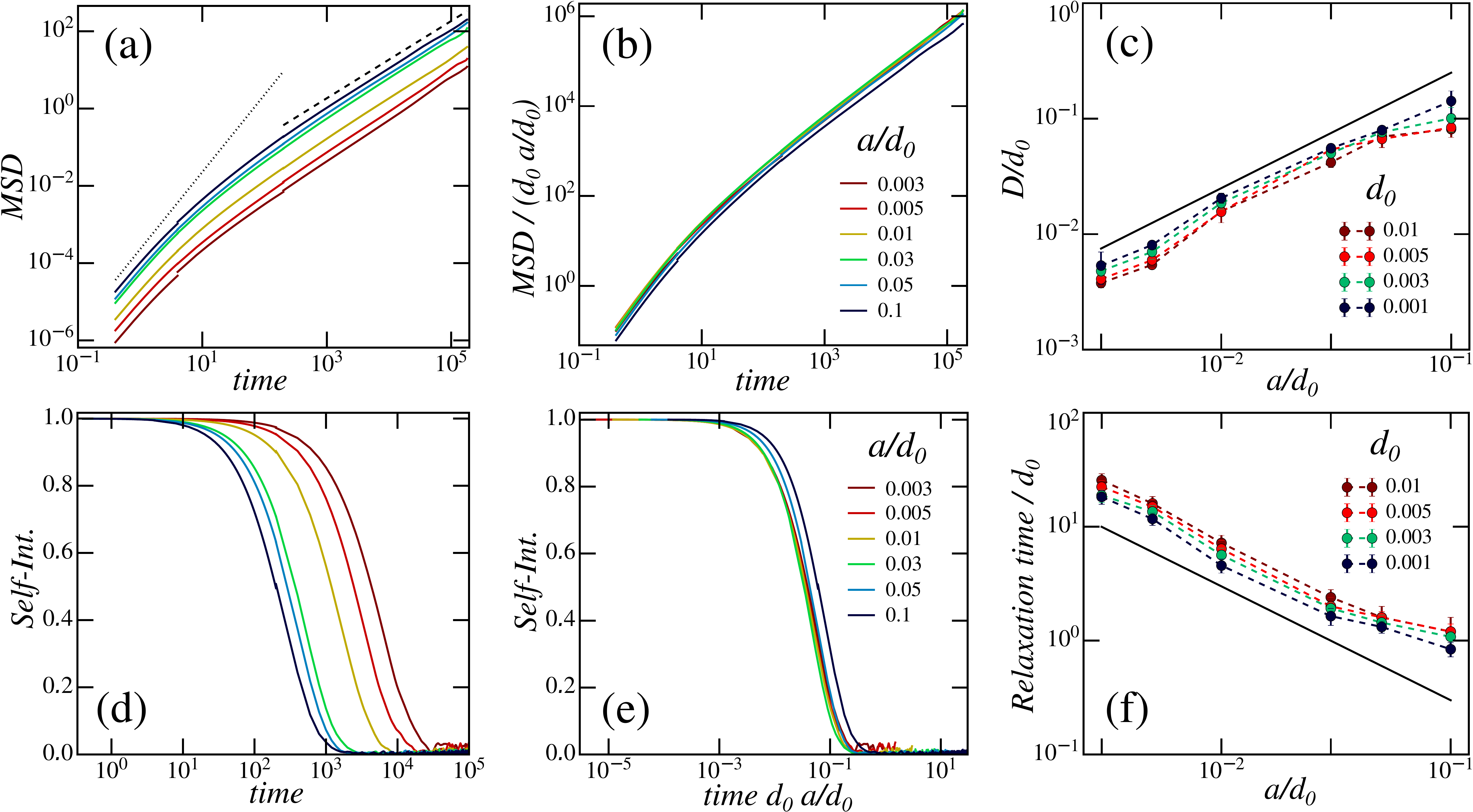}
  \caption{Diffusive dynamics in the dense system. (a) Mean-square 
  displacement for bare division rate $d_0= 0.003$ and a range of 
  $a/d_0$. (b) Rescaling with $1/a$ time scale. (c) Scaling of the diffusion coefficient extracted from (a). (d) Self-Intermediate scattering function $F(t)$ for the same parameters as (a). (e) Rescaling with $1/a$ time scale.  (f) Scaling of the fluidization time scale $\tau$ extracted from (b).}
\label{fig:divide_dense}
\end{figure*}

\subsection{Self-melting confluent tissue}
\label{Self-melting}
The key result here concerns the confluent tissue, where we observe a very slow dynamics, which is nevertheless fluidised by the presence of dividing and dying particles. We illustrate the dynamics of this state in Fig. \ref{fig:2}(c),
where particles are colour-coded by their velocity magnitude. Each individual death or division event is responsible for a displacement wave that propagates diffusively, and that, together with other events, leads to rearrangements in the system and eventually to a finite diffusive motion of cells. 
This dynamics leads to a liquid state at all values of the activity quantified by $d_0$ and $a$. 
We measure the mean-square displacement ${\text{MSD}(t)=\langle |\mathbf{r}(t)-\mathbf{r}(0)|^2 \rangle}$
{{by tracking cells directly from their birth from a division event until death 
(blue particles in Fig. \ref{fig:2}(c)).
In Figure \ref{fig:msd_cells}, we first show that cells follow a characteristic displacement profile, consisting of a rapid push away from the mother, then a plateau period, and eventually diffusion (see panel b). While at first glance these resemble the MSD curves of a supercooled liquid, the origin of the behaviour is very different. In panel a, it becomes apparent that, regardless of the value of $a/d_0$, more than half the cells reach the cage breaking threshold $MSD\approx 1$. There is also no sign of the gaussian displacement profiles expected from caging: the probability distribution of large displacements is exponential (see inset), which is consistent with a process combining a constant death rate with spatial diffusion.

To avoid the confounding influence of the inital division event, and the poor statistics at large times, we add $2\%$ of non-dividing but otherwise identical tracer particles to the system. Panel (b) shows that while at large times, the tracers follow the same diffusive curves as the cells, their behaviour remains diffusive down to very short length and time scales. We focus on the tracer particles in the following, as they represent the overall tissue flow.
}}

Figure \ref{fig:divide_dense}(a) shows a typical set of MSD curves. We observe a ballistic scaling at short times and very small displacements, characteristic of the persistent motion due to individual division or death events and thus dependent on the internal relaxation dynamics characterised by the surface friction $1/\mu$ and the elastic stiffness $k$ of individual cells (see section D of the SI). The strain field caused by these events
corresponds to classical long range elasticity~\cite{MatozPuosi17} as a response to the changes in local structure. Signatures of this elastic response can be seen in the velocity field in Fig. \ref{fig:2}(c). In the long time limit at times longer than a characteristic time $\tau$, the dynamics become diffusive. 
From the long-time behaviour of the tracer motion we define a diffusion coefficient $D$ from $\text{MSD}(t) = 4 D t$. 
In addition, Fig. \ref{fig:divide_dense}(c) shows the scaled diffusion 
coefficient $D/d_0$ as a function of the division/death ratio $a/d_0$. As can be seen, the curves collapse consistently with a linear scaling $D/d_0 \sim a/d_0$, with some deviations for the largest values of the activity $a/d_0$. This last result is in accordance with the theoretical description presented by Ranft \emph{et}.al ~\cite{Ranft2010}.

To better understand how cells decorrelate their positions in time,
we compute the self-intermediate scattering function ${F(t)=\frac{1}{N} \langle \sum_{n=1}^N e^{i\mathbf{q}\cdot (\mathbf{r}_n(t)-\mathbf{r}_n(0))} \rangle_t}$, for a value of $|\mathbf{q}|=\sqrt{2} \pi/\sigma$. As in ordinary liquids and unlike in glassy or supercooled systems, we find a single decay time scale, as shown in Fig. \ref{fig:divide_dense}(d).
We fit the decorrelation time $\tau$ at which $F(t)$ has decayed by half. As shown in panel (f), we observe a simple scaling collapse, ${\tau d_0 \sim (a/d_0)^{-1}}$ as a very good approximation, with again deviations at the largest $a/d_0$. In panel (e), we have rescaled time by the effective inverse time scale $(a/d_0) d_0 = a$, \emph{i.e.} the apoptosis rate. We observe collapse of the curves, and the same holds for the MSD curves (panel (b)). This means that the only relevant time scale for fluidisation is the division time scale proportional to $1/a$ in the stationary state. {{This fluidisation dynamics is independent of system size for $L\geq 60$ (see section C of the SI), and there is no indication of a phase transition. We also emphasise that in this model at $a=d_0=0$ there is simply no motion whatsoever.}}

\begin{figure*}[t!]
\centering
\includegraphics[width=1.0\textwidth]{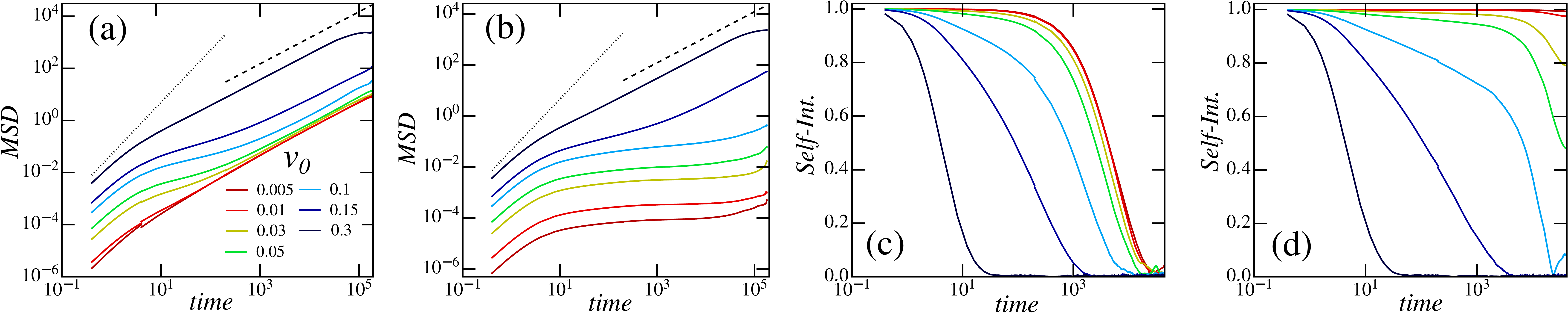}
\caption{Interaction of division/death dynamics with the active glass 
transition induced by self-propulsion. Left: Mean squared displacement. Right: Self-intermediate scattering function. (a) and (c) are for a system with both division/death dynamics and self-propulsion, and (b) and (d) are for a system with only self-propulsion.}
\label{fig:divide_and_active}
\end{figure*}

{{To show how the fluidisation time scale $\tau$ relates to other driving sources in cell sheets}}, we 
 added individual motility to the particles. We use a standard form 
of active dynamics,~\cite{CristinaMarchetti2016} a non-aligning active 
force term ${F_{\text{act}}=v_0 \hat{n}}$, where the unit vector 
$\hat{n}$ diffuses with rotational diffusion coefficient $D_r$. It has been shown that in the absence of division or death, this dynamics leads to a glassy phase at sufficiently high density and low $v_0$.~\cite{Berthier2014,Fily2014} For high values of $D_r$, the system can be mapped to a thermal system with effective temperature ${T_{\text{eff}}=v_0^2/2 D_r}$ and mostly analogous glassy dynamics.~\cite{Berthier2016} Here we consider the case of $D_r =1$, which fits into this regime. In Fig. \ref{fig:divide_and_active} we compare the system with only active motion (panels (b) and (d)) to a system with both active motion and a very small rate of division and death (panels (a) and (c)).

In the system with only active driving, we see a clear transition through the active glass transition as a function of $v_0$. 
The MSD (panel (b)) shows an indefinite plateau at low $v_0$ 
which then increases quadratically with $v_0$, until it reaches the cage breaking threshold.
Panel (d) shows the self-intermediate scattering function characterising the decorrelation of cell positions. As expected for a system with glassy dynamics, $F(t)$ does not decay significantly for the low $v_0$ systems, but decays at increasingly shorter time scales for larger $v_0$. The actual shape of $F(t)$ 
exhibits a stretched exponential decay visible over the whole time 
range. This is likely due to the active nature of the dynamics, and the known effects of $d=2$ on the detailed phenomenology of the glass transition.~\cite{Flenner2015}

If we now add a small amount of cell division dynamics ($d_0= 3\times10^{-3}$ and $a/d_0= 3\times10^{-3}$), we observe that the active dynamics of the system is fully dominated by cell division/apoptosis events. This leads to a complete decorrelation of the positions, \emph{i.e.} a fluidised tissue (panel (c)), and purely diffusive dynamics of the $\text{MSD}$ beyond the ballistic time scale (panel (a)). 
The decay of the intermediate structure factor $F(t)$ (panel (c)) for dynamics with cell division is unaffected only for the largest $v_0$, with a decay that is otherwise truncated by the rapid decay of the dividing contribution. In the same way, only the MSD for the largest $v_0$ that was already diffusive without the division is unaffected. The curves at low $v_0$ essentially collapse on top of the division-only curve.
This remarkable results demonstrate that at long time scales, the division dynamic dominates for low values of the driving $v_0$, therefore erasing any signatures of the glassy state.

\subsection{Gel phase}
\label{sec:gel}
In the intermediate activity rate region, above the percolation point, 
we observe either a confluent tissue, or a phase separated system with strong density heterogeneities. This gel phase is absent at low adhesion strengths ($\epsilon\leq 0.05$) and dominates at larger adhesion values. In order to quantify this gel phase we analyse the coarse-grained density field.~\cite{testard2014} First, we discretise space into boxes of length $\xi_{b}$ and define a discrete density field $\rho(\mathbf{r})$ for discrete positions $\mathbf{r}$ located at the centre of the boxes
\begin{equation}
\label{eq:4}
	\rho(\mathbf{r}) = \frac{1}{V_{b}} \sum_{i}\theta(b - \vert \mathbf{r} - \mathbf{r}_{i} \vert),
\end{equation}
where $V_{b}=\xi_b^2$ is the elementary volume, $\theta(x)$ is the Heaviside function and $b=\langle b_i \rangle$ is the mean particle radius. The coarse grained density field $\bar{\rho}(\mathbf{r})$ is smoothed over adjacent boxes:
\begin{equation}
\label{eq:5}
	\bar{\rho}(\mathbf{r}) = \frac{1}{6} \left[ 2\,\rho(\mathbf{r}) +\sum_{\pm} \sum_{\alpha=x,y} \rho(\mathbf{r} \pm b\: \mathbf{e}_{\alpha}) \right],
\end{equation}
where $\mathbf{e}_{\alpha}$ is the unit vector in the $\alpha$ direction. As in Testard \emph{et}. al ~\cite{testard2014} we set $\xi_{b} = 0.5 b$.

\begin{figure}[h!]
\centering
\includegraphics[width=0.4\textwidth,keepaspectratio]{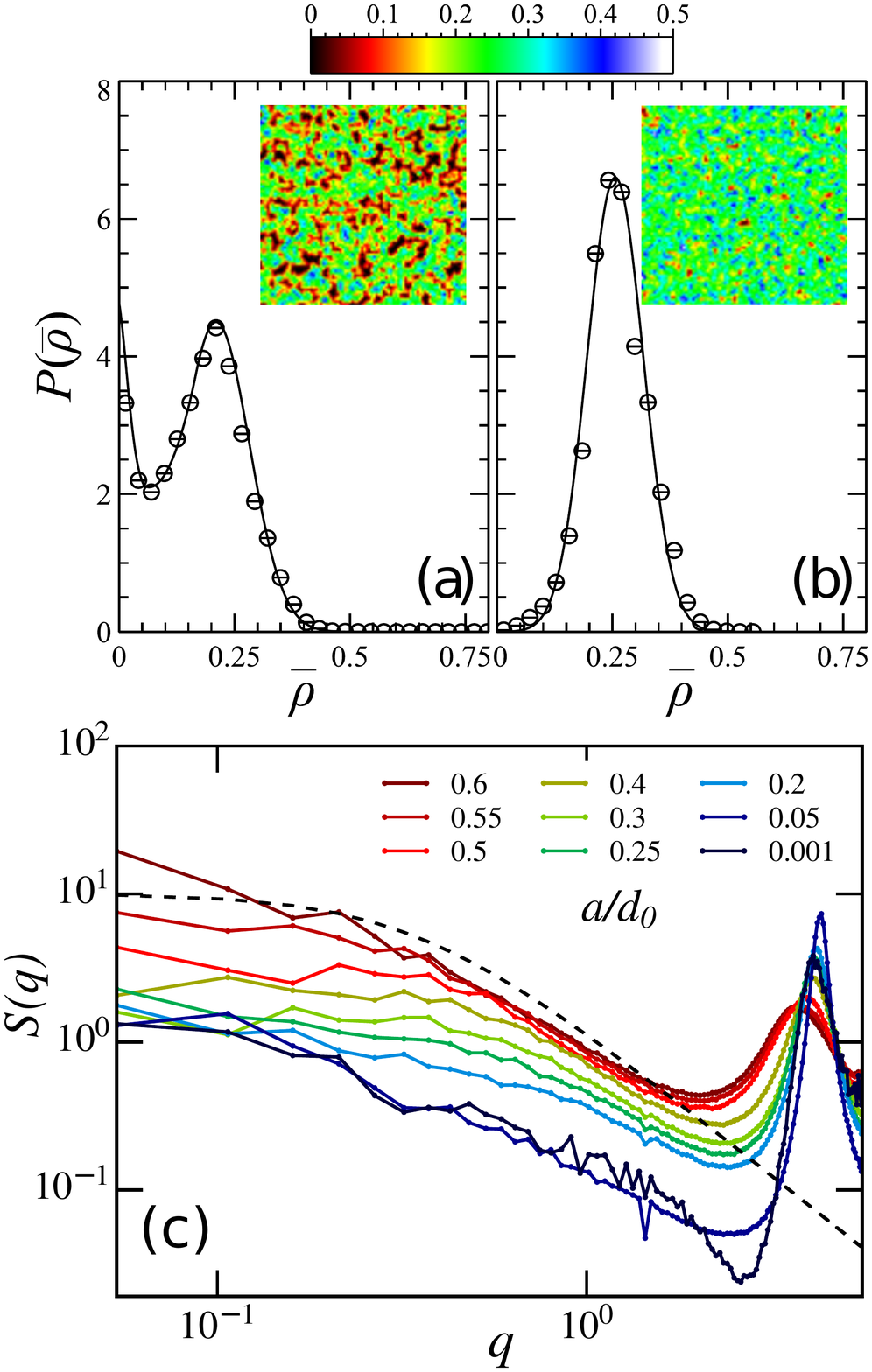}
\caption{Probability distribution corresponding to the coarse-grained density field, Eq. (\ref{eq:5}) for $\epsilon = 1.15$, $d_0 = 10^{-3}$ and $a/d_0$ ratio equal to (a) $2\times 10^{-1}$ and (b) $10^{-3}$. {{(c) Structure factor for the same $\epsilon = 1.15$, $d_0 = 10^{-3}$ for $a/d_0$ decreasing from the non-percolated to the gel and the confluent phase. The dashed line is $\sim 1/(1+\xi^2 q^{2})$, $\xi^2=8$.}}  }
 \label{fig:6}
\end{figure}

In Fig.~\ref{fig:6}, we show typical density fields for $\epsilon = 1.15$ and two $a/d_0$ rates on both sides of the transition. As can be seen from Fig.~\ref{fig:6}(a), for $a/d_0=0.2$, the system is in a phase coexistence state characterised by a strongly heterogeneous coarse-grained density. On the other hand, for a very low apoptosis rate $a/d_0=10^{-3}$ (Fig. \ref{fig:6}(b)), the system is homogeneous.
The probability distribution of the coarse-grained density $P(\bar{\rho})$ gives us a systematic method to distinguish between the intermediate gel phase and the high density self-melting confluent tissue phase. As can be seen in Fig.~\ref{fig:6}, in the gel phase $P(\bar{\rho})$ is characterised by two peaks reflecting phase coexistence. One peak is located at almost zero density representing the non-percolated phase (gas phase). A second peak is at intermediate density representing the cluster phase 
On the other hand, for the high density self-melting confluent tissue the probability distribution of the coarse-grained density is represented by a single peak as expected. We used the presence of a second peak to construct the \textit{Phase-separated - Self-melting} transition line showed in Fig.~\ref{fig:5}. 
{{In Fig.~\ref{fig:6}c, we show the static structure factor $S(q) = \frac{1}{N}|\sum_{n=1}^N e^{i\mathbf{q} \cdot \mathbf{r}_n}|^2$ for a cut through the phase diagram varying $a/d_0$. In addition to a peak corresponding to the position of the nearest neighbours at all $a/d_0$, we find an increase of $S(q\rightarrow 0$). This peak increases with decreasing $a/d_0$ right up to the absorbing boundary, and is consistent with a scaling $S(q) \sim (1+\xi^2 q^2)^{-1}$, with increasing $\xi$, cut off by the system size. This result agrees with the measured fractal dimension $D=2$ of the percolation clusters, and we conclude that the gel phase clusters also have $D=2$. }}

Interestingly, the formation of the gel is accompanied by the build-up of a negative pressure in the system as demonstrated in Fig.~\ref{fig:4}. The source of this negative pressure is that the percolated network structure may exhibit tensile stresses due to the attractive forces when confined to a fixed volume. Therefore, the measurement of a global quantity like the pressure can already give some information on the underlying internal structure. 

\section{Conclusions}

In summary, using a simple model that includes only three independent parameters, we have been able to explore active dynamics relevant to tissues. As we increase the apoptosis rate for a given adhesion force, we encounter, sequentially, a dense confluent tissue phase, a network forming phase, a low density clustering phase and a region where the tissue is dying. We observe that in confluent tissues, regardless of the level of active driving our model fluidises at long times, above the division time scale. Signatures of active glassy dynamics only exist at very short time scales, however they are already severely affected by the division dynamics. We emphasise that this behaviour is not solely a property of the model presented here. For example, in an active vertex model simulation,~\cite{SamosPaper} we have confirmed that adding cell division as only source of activity also fluidises the tissue. {{We have also carried out a full rheological analysis of this model,~\cite{rheology_daniel16} and confirmed the fluid behaviour.}}

The absence of a glassy phase in a system with any level of division or death events is important for the biology of tissues. Our results suggests that in actual developmental epithelial tissues (\emph{e.g.} drosophila, chick embryo and the mammalian cornea), where there is substantial division dynamics, active glassy dynamics does \emph{not} play a fundamental role.  Only \emph{in vitro} systems that have suppressed division rates are more likely candidates to show true glassy features. A number of recent results~\cite{Park2015,Bi2016} predict a glassy phase in confluent tissues, based on a shape parameter relating perimeter and area of cells. However, the associated models~\cite{Bi2016,farhadifar2007influence} all neglect cell division and death.

In further studies it will be important to also consider other 
biological processes that involve more complex collective processes. During organ development or tumour growth, the cells organise themselves in a collective manner by regulating proliferation rate (cell division) and cell death (apoptosis). Gene expression and tissue pattern formation can be highly influenced by the spatial distribution of mechanical stresses.~\cite{Farge2003, Fernandez-Gonzalez2009}

\section*{Acknowledgements}
J.-L.~B. and D.~A.~M.-F.~acknowledge financial support from ERC 
grant ADG20110209. 
J.-L.~B., D.~A.~M.-F. and K.~M. thanks the NVIDIA Corporation for the hardware grant through the Academic Partnership Program.
SH and RS would like to thank Prof Inke N\"athke and Prof Kees Weijer for many illuminated discussion on biology of developing tissues. RS acknowledges  support by the UK EPSRC (award EP/M009599/1) and BBSRC (award BB/N009789/1). SH acknowledges support by the CPTGA visiting researcher fund that allowed her to spend time in Grenoble and the BBSRC (award BB/N009150/1).

\end{document}